# Self-stabilized charge states in a double-decker molecular magnet on Pb(111)

Xin Liao[1,2#], Rui-Jing Sun[1,2#], Emi Minamitani[3], Lian-Zhi Yang[1,2], Tao Xie[1,2], Wen-Hao Zhang[1,2], Chaofei Liu[1,2], Svetlana Klyatskaya[4], Mario Ruben[4,5,6], and Ying-Shuang Fu[1,2*]

1. School of Physics and Wuhan National High Magnetic Field Center, Huazhong University of Science and Technology, Wuhan 430074, China
2. Wuhan Institute of Quantum Technology, Wuhan 430206, China
3. SANKEN, Osaka University, Ibaraki, Osaka, 567-0047, Japan
4. Institute of Nanotechnology, Karlsruhe Institute of Technology, Hermann-von-Helmholtz-Platz 1, 76344, Eggenstein-Leopoldshafen, Germany
5. Institute for Quantum Materials and Technologies, Karlsruhe Institute of Technology, Hermann-von-Helmholtz-Platz 1, 76344, Eggenstein-Leopoldshafen, Germany
6. Centre Européen de Sciences Quantiques, Institut de Science et d'Ingénierie Supramoléculaires, 8 allée Gaspard Monge, BP 70028, 67083 Strasbourg Cedex, France

[#]These authors contribute equally to this work.
*Email: yfu@hust.edu.cn

**ABSTRACT:**
**Electron charging play key roles in physiochemical processes, whose intrinsic stabilization in single molecules is desirable for tailoring molecular functionality and developing molecular devices, but remains elusive on surfaces. Here, we show molecular charge states can be self-stabilized via intramolecular distortion in single bis(phthalocyaninato)terbium(III) (TbPc$_2$) double-decker molecules, that were grown on Pb(111) substrate. Using scanning tunneling microscopy and**

**spectroscopy, we identify fractions of TbPc$_2$ molecules reduce to 2-fold symmetry, expressing energy-split molecular orbitals and two types of different spin states. Our first principles calculations unveil that the symmetry reduction is induced by charging-triggered Jahn-Teller distortions, which lifts the degenerate orbitals into two 2-fold symmetric orbitals. Single or double occupancy of the lower-energy orbital results in different molecular spin states. Such intramolecular distortion traps the excess electrons stably without explicit involvement of the substrate, in contrast to previously observed molecular charge states. These charged single molecule can be manipulated with the tip individually. This study offers a new avenue for tailoring the charge and spin states of molecules.**

**INTRODUCTION:**

Excess charges localized on atomic or molecular solids act crucially in a diverse range of physicochemical processes. They significantly influence charge transport [1,2], surface reactivity [3,4], catalytic activity [5,6], thermoelectric and multiferroic functionality [7,8], etc. Understanding those processes ideally requires the capability of controlling charge localization at single atom or molecule level. Single molecules offer a unique platform for exploring charge-related phenomena due to their versatile functions and the high degree of freedom in chemical modification. Controlling the charge states of molecules provides a powerful means of tailoring their physical and chemical properties. The charging and discharging of electrons would modify the structure and spin states of the molecules, which properties are used for building functional molecular devices, such as molecular spin switches [9-12].

Scanning tunneling microscopy and spectroscopy (STM/STS) have capabilities of characterizing the morphology and electronic states at atomic resolution, serving as an ideal tool for investigating the molecular charge states. In previous studies, numerous molecular systems have been successfully demonstrated to host stable charge states [12-22]. However, a common limitation is that these molecular charge states often rely on the supporting substrates for stabilization. On insulating substrates, the atomic structures of the substrates usually undergo polaronic relaxations to respond and stabilize the excess charges on the molecules [12-18]. On conductive substrates, molecules get charged via substrate charge transfer, as is enforced and stabilized by the chemical potential difference at the molecule/substrate interfaces [19-22]. Consequently, such charge manipulation procedures are highly specific to molecule/substrate combinations and lack universality. Self-stabilized molecular charge states necessitate internal molecular relaxations to trap the excess charges, thereby exempting from the constraints of the substrates. While charged molecules are easily obtained in the gas phase, their charge states usually change on surfaces, as influenced by the local chemical environment. This makes the intrinsically charged molecules on surfaces remain out of reach.

In this work, we report the observation of intrinsically stabilized charge localization in bis(phthalocyaninato)terbium(III) (TbPc$_2$) double-decker molecules, that self-assemble into monolayer films on Pb(111) surface. Inside the molecular films, 2-fold symmetric TbPc$_2$ are observed coexisting with the common 4-fold symmetric TbPc$_2$. Such symmetry reduction of TbPc$_2$, as unveiled cooperatively from molecular orbital mapping and spin state characterization, originates from charging of the molecule, which causes an intramolecular Jahn-Teller (JT) distortion and stabilizes the excess electrons residing on the molecule. As is substantiated with first principles calculations, such charges stabilization is intrinsic, without assistance from the substrate. The single charged-molecule can be steadily moved at surfaces with tip manipulation, and releases its excess electrons via high tip electric fields. This study discovers self-stabilized molecular charge states, paving the way for their application in building miniaturized molecular devices.

**EXPERIMENTAL SECTION:**

The measurements were performed in a custom-made Unisoku STM system (1300) at 0.4 K under ultra-high vacuum conditions. $TbPc_2$ molecules were thermally evaporated from a home-built evaporator at 650–700 K, where accessory tris(phthalocyaninato)di(terbium(III)) ($Tb_2Pc_3$) are formed via a chemical reaction from $TbPc_2$ precursors [23-25]. The molecules were deposited on large Pb(111) film, which were pre-grown on $SrTiO_3$(001) substrate by molecular beam epitaxy (MBE). The default substrate temperature is 300 K during molecule deposition, unless stated otherwise. The first principles calculation is carried out by density functional theory (DFT) with Vienna ab-initio simulation (VASP) package [26-28] with the projected augmented wave method. The correction via van der Waals interaction developed by Hamada [29] was used. The detailed methods are depicted in the supplementary information.

**RESULSTS AND DISCUSSION:**

The $TbPc_2$ molecule consists of one Tb ions sandwiched by two Pc ligands that mutually stack by 45º [insert of Fig. 1(a)]. This peculiar molecular structure has two merits: (1) The lower Pc largely decouples the upper Pc from strong coupling with the substrate, rendering the molecular orbital features well reserved. (2) The molecule possesses flexible internal degree of freedom for structural distortion, which is a prerequisite for stabilizing the excess charges via structural relaxations.

STM image unveils they self-assemble into square lattices upon adsorption on Pb(111) surface [Fig. 1(a)]. Several $Tb_2Pc_3$ molecules, that appear brighter, are imbedded inside the $TbPc_2$ molecular lattice [24, 25]. The zoom-in image clearly visualize the 4-fold symmetric $TbPc_2$ molecules with eight-lobed structures [Fig. 1(b)], dubbed as C4-$TbPc_2$, corresponding to their top Pc ligands. Notably, a number of $TbPc_2$ molecules apparently possess 2-fold symmetry [Fig. 1(b), dashed circles], named as C2-$TbPc_2$, where two pairs of lobes appear brighter that the other two pairs. The brighter lobes are distributed on two-opposite lobes of lower Pc, as is seen from the comparison between the superimposed structural model of the top Pc and the STM image of the C2-$TbPc_2$ [Fig. 1(b)]. Isolated $TbPc_2$ has an unpaired π-electron spin delocalized over its two Pc ligands and a f-electron spin on the Tb ion, acting as a single molecule magnet with high blocking temperature [30, 31]. Extensive interests have been sparkled on the investigation of $TbPc_2$ molecules on surfaces [13, 31-39], particularly with STM. Nevertheless, the f-spin is usually too localized to be detected [24,25] and thus not discussed hereafter. Upon adsorption on Pb, the ligand of $TbPc_2$ readily accepts an electron transferred from the substrate [24,25], making its π-electron spin paired up and becomes nonmagnetic, similar to the cases on Ag(111) [23] and Cu(111) [39] substrate. The molecular anion $[TbPc_2]^-$ remains 4-fold symmetric, as reported previously [24,39].

Nevertheless, the TbPc$_2$ molecule can be charged with more than one electrons. As will be shown below, the symmetry reduction of the molecule originates from its multi-valence charge states.

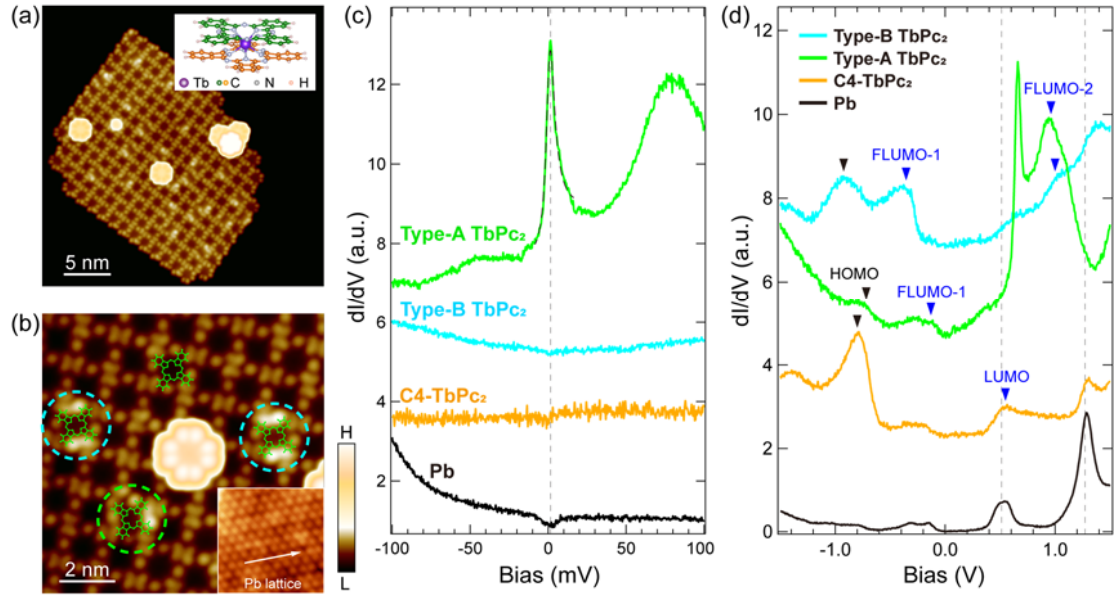

Figure 1. Topography and spectra of TbPc$_2$ molecules. (a) STM images ($V$ = -1.0 V and $I$ = 10 pA) of a self-assembled TbPc$_2$ molecular film on Pb(111). The brighter molecules are Tb$_2$Pc$_3$. Structural model of TbPc$_2$ is shown in the inset. (b) Enlarged STM image showing the C2-TbPc$_2$ molecules embedded inside the molecular lattice coexisting with the C4-TbPc$_2$ molecules. Green and cyan dashed circles mark typical Types-A and -B TbPc$_2$, respectively. Structural model of the top Pc is superimposed. Inset in (b) shows an atomic resolution image ($V$ = -5.0 mV and $I$ = 30 nA) of the Pb(111) substrate, where one of the Pb lattice directions is marked with a white arrow. (c) Low energy d$I$/d$V$ spectra ($V$ = -100 mV and $I$ = 200 pA. $V_{\text{mod}}$ = 1 mV) taken on the C4-TbPc$_2$, Types-A and -B TbPc$_2$ molecules under B = 2 T. The dashed curve on the spectrum of Type-A TbPc$_2$ is a fit to Kondo peak. (d) Tunneling spectra ($V$ = -1.5 V and $I$ = 100 pA. $V_{\text{mod}}$ = 20 mV) of the different types of TbPc$_2$ molecules and the Pb substrate. The dashed lines mark the characteristic quantum well states of the Pb(111) substrate. Characteristic molecular orbital energies are marked with colored triangles. The spectra in (c,d) were all acquired at the molecular centers.

Charge transfer from the substrate induces spin state transition of the molecule, which in turn can be utilized to characterize the charge state [9]. To examine the spin state, we measured low-energy spin excitation spectra of the molecule, which manifest Kondo resonances around the Fermi level $E_F$ due to the spin screening of the molecular spin by the itinerant electrons of the substrate. As shown in Fig. 1(c), spectrum of the C4-TbPc$_2$ molecule is featureless, demonstrating its ligand spin is inactive and conforming to the anionic state of [TbPc$_2$]$^-$. Interestingly, spectra of the C2-TbPc$_2$ molecules display two different types, named as Types-A and -B, respectively. The Type-A TbPc$_2$ exhibits a sharp peak at $E_F$, which is ascribed to a Kondo resonance in view of its narrow peak width of 3.30 meV [Figs. 1(c) and S1]. The emergence of Kondo resonance suggests a different molecular charge state compared to [TbPc$_2$]$^-$. Since both [TbPc$_2$]$^0$ and [TbPc$_2$]$^-$ are 4-fold symmetric, the Type-A TbPc$_2$ should possess a distinct charge state, whose most plausible valence state could be [TbPc$_2$]$^{2-}$. Namely, an excess electron may transfer onto the lowest unoccupied molecular orbital (LUMO) of [TbPc$_2$]$^-$, making it singly occupied and thus spin active. By contrast, the Type-B TbPc$_2$ displays featureless spectrum again, implying its charge states is different from that of the Type-A TbPc$_2$, despite of their similar 2-fold symmetric morphology. Statistics over many molecular films indicate about 1.5% of the TbPc$_2$ molecules are 2-fold symmetric, wherein the population ratio of Types-A and -B molecules is about 1:3.

Having identified the spin states of the molecules, we characterize their spectroscopic features in large energy scale [Fig. 1(d)]. The spectrum of C4-TbPc$_2$ shows two molecular states at -0.80 and 0.54 eV, corresponding to the highest occupied molecular orbital (HOMO) and LUMO respectively of [TbPc$_2$]$^-$ [24, 39]. There are two additional peaks at 0.52 and 1.28 eV on the molecule, which are however extrinsic, because they come from the wave function tails of the quantum well states of the Pb substrate [40]. The spectrum of Type-A TbPc$_2$ [Fig. 1(d), green curve] shows two negative-bias peaks at -0.72 and -0.12 eV, as well as two positive-bias peaks at 0.65 and 1.00 eV. The peak at 0.65 eV is significantly sharper than the other peaks, whose energy not only varies prominently on different Type-A TbPc$_2$ molecules, but also shifts progressively towards higher energy as the tip-sample distance decreases (Fig. S2). This demonstrates the sharp peak is essentially the same state as the peak at -0.12 eV, which are detected at both positive and negative bias, due to the electron transport in the double barrier tunneling junction composed of tip-molecule and molecule-substrate barrier [13, 41]. The Type-B TbPc$_2$ shares some similar spectroscopic features to the Type-A molecule but with notable differences as well. For one thing, the Type-B TbPc$_2$ also has two occupied states, but at lower energies of -0.90 and -0.35 eV. For another, there is only one unoccupied state at 1.00 eV. Both Types-A and -B TbPc$_2$ have small peaks around 0.52 eV and 1.28 eV, which are again the emanating quantum well states of the Pb substrate.

To unravel the molecular states in the spectra of the C2-TbPc$_2$, we perform spectroscopic mapping on the relevant molecular state energies (Fig. 2). The HOMO and LUMO states of the C4-TbPc$_2$ all indicate 4-fold symmetry (Fig. S3), conforming to those states expected for [TbPc$_2$]$^-$ [46]. In contrast, spectroscopic mapping of Type-A TbPc$_2$ unveils the molecular state at -0.12 eV possesses a 2-fold symmetry [Fig. 2(b)], which is oriented along the same direction as the bright lobe pair of the molecular morphology [Fig. 2(a)]. Interestingly, the molecular state at 1.00 eV also has a 2-fold symmetric distribution [Fig. 2(d)], but is oriented orthogonal to that of the -0.12 eV. The observed two orthogonal molecular states are reminiscent of an excess electron charging onto a doubly degenerated LUMO orbital of [TbPc$_2$]$^-$ as calculated previously [46], which induces JT distortion to the molecule and lifts the degeneracy of the LUMO [22], as will be shown in Fig. 4(a). As such, the Type-A TbPc$_2$ is ascribed to a charge state of [TbPc$_2$]$^{2-}$, and the two molecular states at -0.12 eV and -1.00 eV are labeled as FLUMO-1 and FLUMO-2, respectively. The FLUMO state means the former LUMO state of [TbPc$_2$]$^-$ after one additional electron charging. The Kondo resonance mapping coincides with that of the FLUMO-1 [Fig. 2(c)], further substantiating the FLUMO-1 state is singly occupied. Similarly, for the Type-B TbPc$_2$ [Fig. 2(e)], its molecular states at -0.35 eV and 1.00 eV are also orthogonal to each other [Figs. 2(f,h)], and thus are ascribed as FLUMO-1 and FLUMO-2, respectively. However, it has no Kondo resonance [Fig. 2(g)], implying its FLUMO-1 state becomes doubly occupied [Fig. 4(a)]. As such, we attribute the charge state of Type-B TbPc$_2$ as [TbPc$_2$]$^{3-}$. Such assumptions will be justified from our DFT calculations later.

Having inferred the charge states of the 2-fold symmetric molecules, we subsequently investigate the mechanism that stabilizes the excess charges. Previous investigations on the charged molecule were supported on thin insulating films, wherein the charge states were stabilized by structural relaxation of the supporting substrate [24].

In this regard, despite our supporting substrate is metallic, we first examine whether the molecular charge states are related to the Pb substrate. Scrutiny over numerous molecular films indicates that the bright lobes of the C2-TbPc$_2$ are approximately equally oriented along the <1 1 0> and <-2 1 1> directions of the Pb(111) substrate, with a statistical ratio of 1:1.07. This demonstrates that C2-TbPc$_2$ molecules have negligible preference in orientation relative to the substrate, suggesting the substrate may have no structural relaxation to stabilize the excess charge.

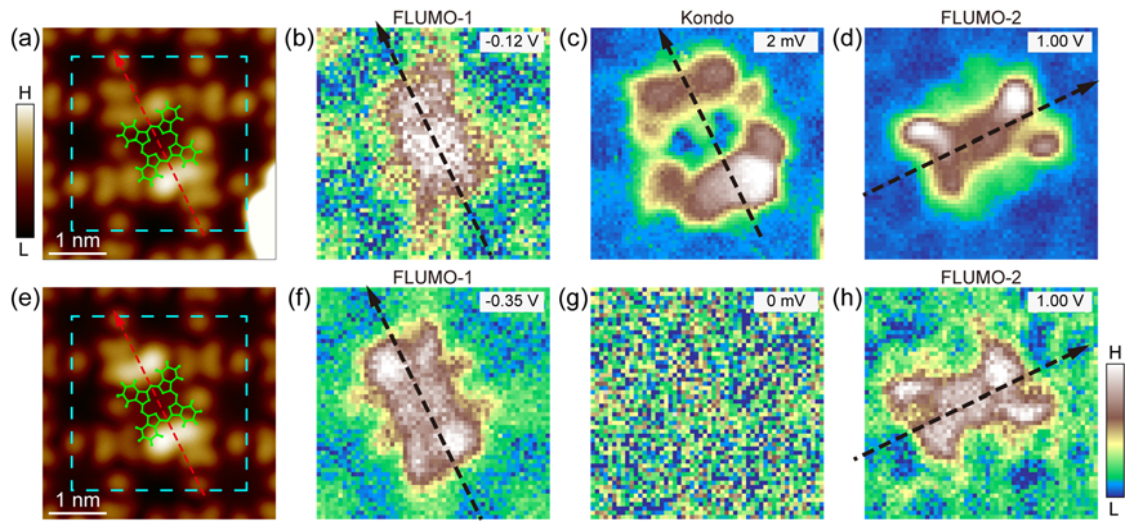

Figure 2. Molecular orbital imaging of C2-TbPc$_2$. (a) High-resolution STM images ($V$ = -1.0 V and $I$ = 10 pA) of Type-A TbPc$_2$. Structural models of top Pc are superimposed. (b-d) d$I$/d$V$ mapping of the rectangle area in (a), taken at the indicated molecular orbital energies (b,d) and the Kondo resonance energy (c), respectively. (e-h) Similar as (a-d), but for the Type-B TbPc$_2$. The mapping in (g) shows the absence of Kondo resonance. The dashed arrows mark the orientation of the higher intensity contrast direction. Spectroscopic conditions: $V$ = -1.5 V and $I$ = 100 pA. $V_{mod}$ = 20 mV for (b,d,f,h); $V$ = -100 mV and $I$ = 200 pA. $V_{mod}$ = 2 mV for (c,g).

We further evaluate whether molecular charge states are stabilized by intermolecular interaction. For that, we manipulated the 2-fold symmetric TbPc$_2$ out of the molecular film with the STM tip. As shown in Fig. 3, both the Types-A and -B TbPc$_2$ maintain their 2-fold symmetry as single molecules isolated from the molecular film. Spectroscopic mappings of the isolated Types-A and -B TbPc$_2$ both show orthogonal spatial distributions between their FLUMO-1 and FLUMO-2 states (Fig. S4), in the same fashion as those in the molecular films. Moreover, the spin states of Types-A and -B TbPc$_2$ are also reserved during the tip manipulation, where the Kondo resonance mapping of the Type-A molecule displays same distribution as the FLUMO-1 state (Fig. S4). All the above observations rigorously exclude the role of intermolecular interaction or the substrate in stabilizing the charge states. As such, the charge states may probably be stabilized by the intrinsic intramolecular structure distortion.

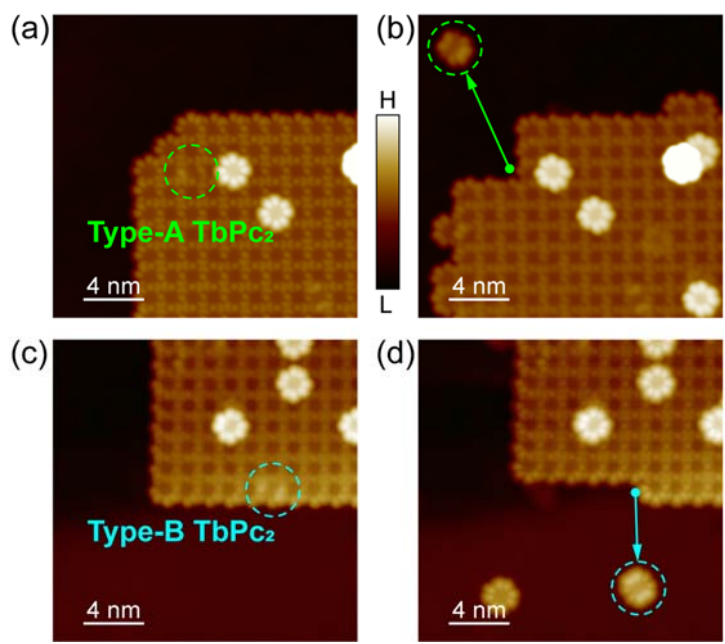

Figure 3. Manipulation of C2-TbPc$_2$. (a, b) STM images ($V = -1.0$ V and $I = 10$ pA) of the embedded Type-A TbPc$_2$ before (a) and after (b) being moved out of the molecular film. (c, d) Similar as (a,b), but for manipulating the Type-B TbPc$_2$. The dashed green (cyan) circle marks the Type-A (Type-B) TbPc$_2$. The arrows mark the manipulation directions.

To unravel specific mechanism of the molecular charge states and spin states, we carried out density functional theory (DFT) calculations to the isolated TbPc$_2$ molecule. Including the substrate would drastically increase the load of the calculation, making it infeasible. We start from calculating the case of C4-TbPc$_2$, where an additional electron is added to become the [TbPc$_2$]$^-$ state. The calculation shows the [TbPc$_2$]$^-$ becomes nonmagnetic, whose HOMO and LUMO states are both 4-fold symmetric (Fig. S5). Particularly, its LUMO state is doubly degenerated [Figs. S5 and 4(a)]. Their density of states are distributed mainly along two opposite lobes and are orthogonal to each other, in agreement with previous report [42].

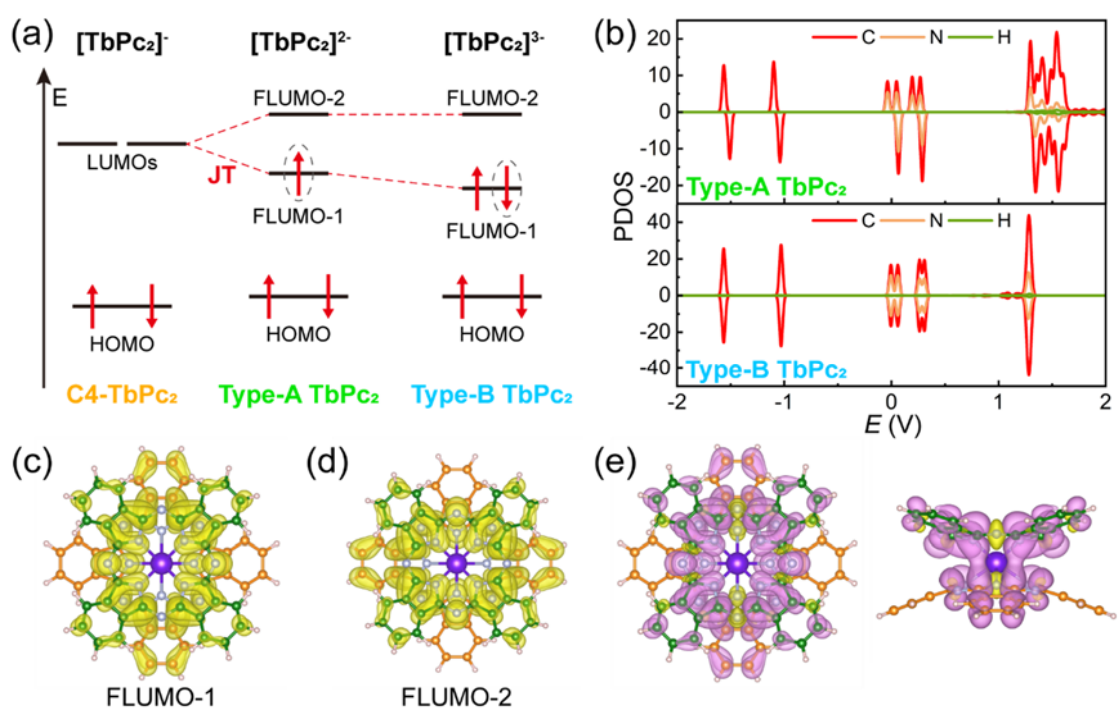

Figure 4. DFT calculations of C2-TbPc$_2$. (a) Schematic energy scheme of the C4-TbPc$_2$, Type-A and Type-B TbPc$_2$ molecules. (b) Projected density of states of the Type-A ([TbPc$_2$]$^{2-}$) and Type-B TbPc$_2$ ([TbPc$_2$]$^{3-}$) after structural optimization. (c, d) Charge distributions of the FLUMO-1 and FLUMO-2 orbital of Type-A TbPc$_2$. (e) Top (left) and front (right) view of the spin density distribution for the majority (red) and minority (yellow) spin components of Type-A TbPc$_2$.

For Type-A TbPc$_2$, we conjectured it has a charge state of [TbPc$_2$]$^{2-}$, where charging-induced JT distortion lifts the degenerate LUMO into FLUMO-1 and FLUMO-2. Single occupancy of the FLUMO-1 state brings about local moment to the Pc ligand. To justify such hypothesis, we performed structural optimization of the [TbPc$_2$]$^{2-}$ molecule. While the optimized [TbPc$_2$]$^{2-}$ retains 4-fold symmetry, we identified three vibrational modes with imaginary frequencies [Fig. S6(a)] in the vibrational spectrum. Among them, two degenerated vibrational modes $v_1$, $v_2$ depict vibrations along the two opposite lobes of bottom Pc, and one mode $v_3$ corresponds to vibrations perpendicular to the Pc plane [Figs. S6(c-e)]. The emergence of imaginary frequency modes demonstrate the 4-fold symmetric molecular structure is unstable and tends to distort. Based on the experimentally observed 2-fold molecular symmetry, we selected one of the two degenerated modes and created a distorted molecular structure by applying displacements along the vibrational mode to the respective atoms. The optimized structure obtained from this distorted configuration successfully eliminates the imaginary frequency modes, and reduces the molecular symmetry into C2. The calculation shows the JT distortion dominantly occurs to the benzene rings and is slightly asymmetric, as shown in Fig. S6(b), which breaks the degenerate LUMOs state into two orthogonal FLUMO-1 and FLUMO-2 state [Figs. 4(c, d)]. The FLUMO-1 state is singly occupied [Fig. 4(b)] and shares the same spatial distribution as its ligand spin [Fig. 4(e)], whose C2 symmetry mainly comes from the bottom Pc, conforming to the experiments [Figs. 2(a-d)]. The calculated energy of the C2-[TbPc$_2$]$^{2-}$ is 0.949 eV lower than that of the C4-[TbPc$_2$]$^{2-}$, in consistency with the JT distortion mechanism.

As for the Type-B TbPc$_2$ molecule, since its ligand spin is absent, we speculate that its FLUMO-1 state is doubly occupied, resulting in a charge state of [TbPc$_2$]$^{3-}$ [Fig. 4(a)]. To confirm this, we evaluated the electronic state by adding an extra electron to the JT distorted [TbPc$_2$]$^{2-}$. After further optimization, we found that the molecular structure of [TbPc$_2$]$^{3-}$ preserves the C2 symmetry of the JT distorted [TbPc$_2$]$^{2-}$, whose asymmetric distortion gets more severe and expands to other Pc lobes, as shown in Fig. S7(a). The FLUMO-1 becomes doubly occupied with the additional electron, causing the ligand spin to vanish [Fig. 4(b)]. The wave function distributions of the FLUMO-1 and FLUMO-2 states remain orthogonal to each other [Fig. S7(b,c)], in agreement with the experiment [Figs. 2(e-h)]. Furthermore, the calculated C2-[TbPc$_2$]$^{3-}$ is 1.012 eV lower in energy that the C4-[TbPc$_2$]$^{3-}$, again conforming to the JT mechanism. The above calculations suggest that excess electrons can be trapped via a JT distortion in the isolated molecule, without the involvement of the supporting substrate. Thus, this validates the existence of self-stabilized charge states in our C2-TbPc$_2$ molecules.

Finally, we demonstrate the feasibility of switching the C2-TbPc$_2$ to C4-TbPc$_2$ via applying high voltages (Fig. 5). By positioning the tip on the center of a C2-TbPc$_2$, the sample bias is ramped from -2 V to 1.9 V. During such process, the tunneling current jumps at about 1.8 V, which signifies the switching from C2- to C4-TbPc$_2$ (Fig. 5). Such switching occurs to both Types-A and -B TbPc$_2$ with similar threshold voltages (Fig. S8). While the C2-TbPc$_2$ can be transformed to C4-TbPc$_4$ with 100% success rate, the reversed transition has never been achieved. This implies the C2-TbPc$_2$ molecules are in a metastable state. The large voltage generates a tip electric field excreting on the C2-TbPc$_2$ with excess charges. This might cause structural relaxation of the molecules, which could compete with the charging-induced JT distortion. The excess charges are released to transform into the C4-TbPc$_2$ upon the tip-electric-field induced structural relaxation overwhelms the JT distortion. Considering that the C2-TbPc$_2$ molecules naturally exist within the self-assembled molecular films, we speculate that they might get charged already inside the crucible prior to their deposition onto the Pb substrate. Presumably, the molecules are charged by electrons emitted out of the crucible during evaporation. We further find the population ratio of C2-TbPc$_2$ molecules increases at lower substrate temperature (Fig. S9), implying that the excess electrons can hop from the molecules onto the substrate assisted via molecular vibrations.

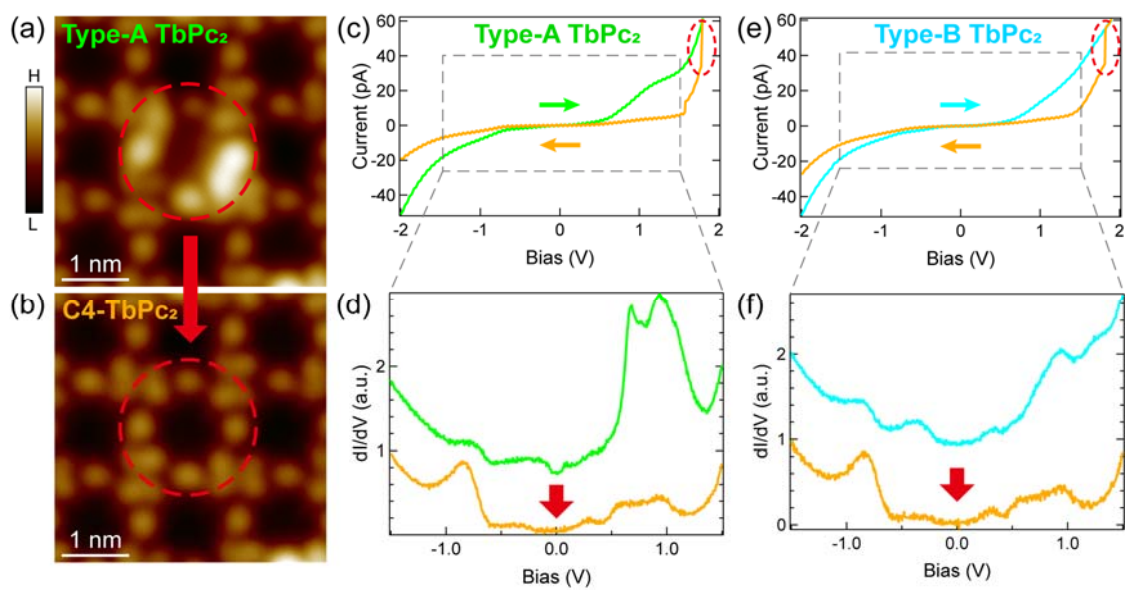

Figure 5. Switching from C2- to C4-TbPc$_2$. (a, b) Typical STM images ($V$ = -1.0 V and $I$ = 10 pA) of a Type-A TbPc$_2$ (a) transformed to a C4- TbPc$_2$ (b). Similar transformation of a Type-B TbPc$_2$ is shown in Fig. S8. Structure model of the top Pc is superimposed on the images. (c,d) Tunneling spectra of Type-ATbPc$_2$, showing the current (c) and d$I$/d$V$ conductance (d) changes (marked with dashed ellipses) upon its switching to C4-TbPc$_2$. (e,f) Similar as (c,d), but for Type-B TbPc$_2$. The arrows mark the voltage

ramping directions. Spectroscopic conductions: $V$ = -1.5 V, $I$ = 100 pA, $V_{mod}$ = 20 mV.

**CONCLUSION:**

In conclusion, we have discovered intrinsically charged TbPc$_2$ molecules, where internal JT distortion of the molecule, rather than the substrate, stabilizes the excess electrons residing on the molecules. Two types of TbPc$_2$ show 2-fold symmetric morphology with different spin states on Pb(111) substrate, which are induced by different charge states of [TbPc$_2$]$^{2-}$ and [TbPc$_2$]$^{3-}$, respectively, in contrast to the commonly observed 4-fold symmetric [TbPc$_2$]$^-$ molecules. Our DFT calculations unveil that the excess electrons result in JT molecular distortions, lifting the otherwise degenerate LUMO states of the [TbPc$_2$]$^-$ into two FLUMO-1 and FLUMO-2 states. Single (double) occupancy of the FLUMO-1 state causes the magnetic (nonmagnetic) spin states of the 2-fold symmetric [TbPc$_2$]$^{2-}$ ([TbPc$_2$]$^{3-}$). Such JT distortion further stabilizes the excess electrons, which can be extended to other molecular species. This offers a new avenue for manipulating the charge and spin states by utilizing the internal structural degree of freedom, which may find its applications such as molecular spintronics, information storage, and quantum information process, etc.

**ACKOWLEGEMENT:**

We acknowledge the financial support from the National Key Research and Development Program of China (Grant No. 2022YFA1402400), the National Natural Science Foundation of China (Grants Nos. 92477137, U20A6002, 92265201, 9247710470, 12174131). MR and SK acknowledge the German Research Foundation (DFG) Collaborative Research Centre (CRC) 1573 "4f for Future".